\documentclass[a4paper,11pt]{article}
\usepackage{lineno}
\pdfoutput=1 
\usepackage{jinstpub} 
\usepackage[colorinlistoftodos]{todonotes}
\usepackage{multirow}
\usepackage{caption}
\usepackage{subcaption}
\usepackage{eqnarray,amsmath}

\usepackage{amssymb}
\usepackage{array}
\usepackage{url}

\title{A measurement of the group velocity of scintillation light in liquid argon }
\author[a,b]{M. Babicz}
\author[a]{,S. Bordoni}
\author[e]{, A. Fava}
\author[a]{, U. Kose}
\author[a]{, M. Nessi}
\author[a, h]{, F. Pietropaolo}
\author[c]{, G.L. Raselli}
\author[a]{, F. Resnati}
\author[c]{, M. Rossella}
\author[a,g]{, P. Sala}
\author[a,i]{, F. Stocker}
\author[a]{, A. Zani}

\emailAdd{stefania.bordoni@cern.ch, umut.kose@cern.ch}

\affiliation[a]{CERN,\\European Organization for Nuclear Reasearch, Geneva, Switzerland}
\affiliation[b]{Institute of Nuclear Physics Polish Academy of Sciences, Cracow, Poland}
\affiliation[c]{INFN Sezione di Pavia,\\Via Bassi 6, 27100 Pavia, Italy}
\affiliation[e]{Fermilab National Laboratory, Chicago, USA}
\affiliation[f]{Department of Physics, University of Pavia,\\Via Bassi 6, 27100 Pavia, Italy}
\affiliation[g]{INFN Sezione di Milano,\\ Via Celoria 16, 20133 Milano, Italy}
\affiliation[h]{INFN Sezione di Padova,\\Via Marzolo 8, 35131 Padova, Italy}
\affiliation[i]{University of Bern - Laboratory of High Energy Physics, \\Bern, Switzerland}

\abstract{
The propagation velocity of scintillation light in liquid argon $v_{g}$ at $\lambda \sim 128$~nm wavelength, has been measured for the first time in a dedicated experimental setup at CERN.\\
The obtained result $\frac{1}{v_{g}} = 7.46 \pm 0.08$~ns/m  , is then used to derive the value of the refractive index (n) and the Rayleigh scattering length ($\mathcal{L}$) for liquid argon in the VUV region. For $\lambda = 128$~nm we found $n= 1.358 \pm 0.003$ and $\mathcal{L}= (99.1 \pm 2.3$)~cm. The measured values are of interest for a variety of experiments searching for rare events like neutrino and dark matter interactions. The derived quantities also represent key information for theoretical models describing the propagation of scintillation light in liquid argon.

}


\begin{document}
\maketitle
\addtocontents{toc}{\protect\setcounter{tocdepth}{2}}
\flushbottom
\section{Introduction}

In the lively field of experimental Particle Physics of rare events, liquefied noble gases are widely used for dark matter searches, studies of the properties of neutrinos and investigations of matter stability. Liquid argon (LAr), in particular, is the top pick for neutrino oscillation experiments, for instance in the Short Baseline Neutrino program at Fermilab \cite{SBN} and in view of the next generation large mass detectors for the long baseline DUNE Project \cite{DUNE}, alongside with dark matter experiments such as DarkSide \cite{DarkSide}. 

Exploitation of scintillation light, possibly combined with the collection of ionisation charge, is the main common point to liquid argon detectors, in order to identify the interacting particles and study their properties.
The precision of these measurements is therefore deeply interlaced with the understanding of liquid argon as a scintillation medium, summarized in Section \ref{sec:SciLight}. 

Mechanisms of production of scintillation light in LAr have been extensively investigated \cite{light_production_exp, LArLight_plot, light_production_exp2, light_production_exp3}, leading to a well established picture. 
On the other hand, phenomena related to the propagation of scintillation photons are either controversial, as for the discrepancy between expected \cite{Rayleigh_th, Rayleigh_th2} and measured \cite{Rayleigh_exp, ArDM, ANeumeier} values of the Rayleigh scattering length, or unexplored, like the refractive index, for which only extrapolations from data at longer wavelengths exist. Bridging this gap is crucial especially in view of future kton scale detectors, in which the distance of the photon-detection system from the interaction point will range up to several meters.

This paper presents a first measurement of the group velocity of scintillation photons in liquid argon. From this measurement we derive the values for the refractive index and the Rayleigh scattering length for the typical wavelength value (128~nm) of the LAr scintillation light. For this purpose a dedicated setup, described in Section \ref{sec:setup}, was prepared at CERN.  
The adopted experimental and analysis strategies lead to a result independent 
of the precision in the time calibration of the instruments, as explained in Section \ref{sec:analysis}.
Section~\ref{sec:finresults} presents the results for the measurement of the inverse of the light velocity in LAr. Furthermore, the section discusses the derivation of important light propagation parameters, namely Rayleigh scattering and the refractive index from the measured quantity.  
A FLUKA simulation 
was developed as a benchmark.

\section{Scintillation light in LAr}
\label{sec:SciLight}

\begin{figure}
 \centering
 \begin{subfigure}{.45\textwidth}
 \centering
 \includegraphics[width=0.9\textwidth]{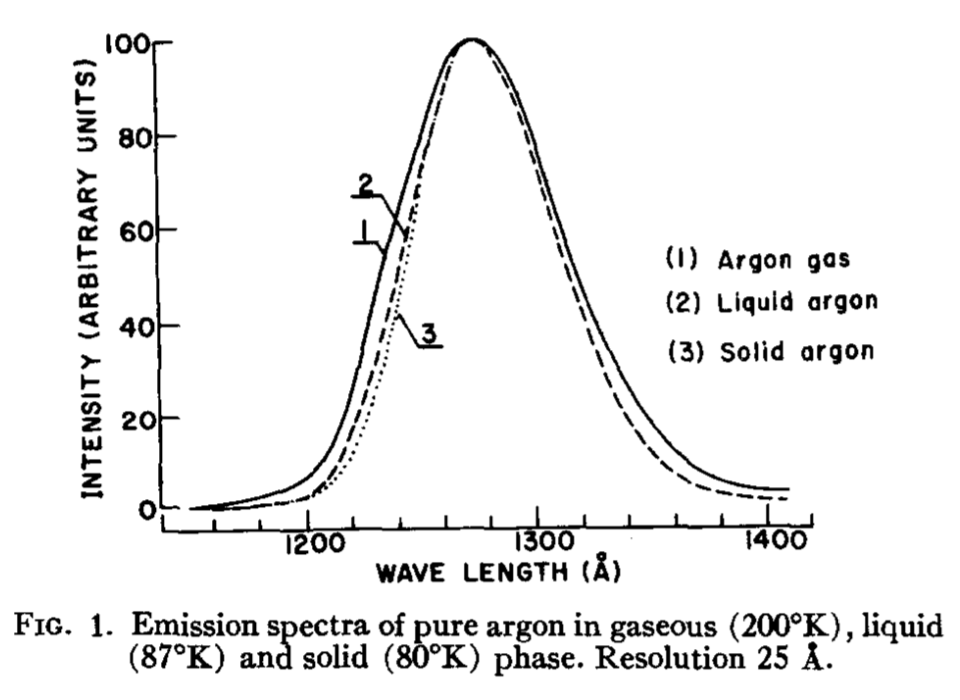} 
 \caption{\label{fig:LArLightspectraA}}
 \end{subfigure}
 \hspace{.2cm}
 \begin{subfigure}{.45\textwidth}
 \centering
 \includegraphics[width=0.9\textwidth]{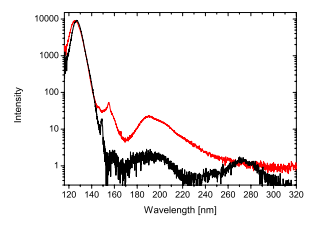} 
\caption{ \label{fig:LArLightspectraB}}
 \end{subfigure}
 \caption{ \label{fig:LArLightspectra} Emission spectra of pure argon from literature. For the liquid phase the distribution is centered at (a) $\lambda = 128$~nm (FWHM =~ 10 nm)~\cite{LArFilippo}, (b) 126 nm (FWHM =~ 7.8 nm) for liquid (black) and gas (red) phases~\cite{light_production_exp}.}
\end{figure}

    The production mechanism of the scintillation light in LAr is known since several decades, literature offers extensive descriptions and its spectrum is well known today. The typical wavelength for the scintillation light from liquid argon is usually cited as $\lambda = 128$~nm~\cite{LArFilippo}. A more recent paper quotes instead $\lambda = 126$~nm~\cite{light_production_exp}. Figure~\ref{fig:LArLightspectra} shows the spectra from these two cited papers. 

In the next decade a number of experiments plan to use LAr for large scale detectors. As mentioned, the optical properties of this medium are not well known yet: controversial measurements or calculations exist for the refractive index or the Rayleigh  scattering length at the LAr emission wavelength. A brief summary of the current knowledge of those two parameters is presented  in Table \ref{table:literature}.

\begin{table}[ht!]
\centering
\caption{Summary of the current knowledge of the optical properties for liquid argon at its scintillation wavelength, from calculations (calc.) and measurements (exp.). The wide range of values for the Rayleigh scattering length is related to the spread in values of $n$, as it will be made clear in Equation~\ref{RayScat} .} 
\label{table:literature}
\begin{tabular}{| c | c | c |}
\hline
 Parameter: & Value: & Measured/calculated by: \\

\hline
Refractive index n           & 1.37                & (calc.) \cite{Rayleigh_th} \\
                             & 1.45 $\pm$ 0.07     & (calc.) \cite{Rayleigh_th2}\\

\hline
Attenuation length (cm)      & 66 $\pm$ 3          & (exp.)  \cite{Rayleigh_exp}\\
                             & 52 $\pm$ 7          & (exp.)  \cite{ArDM}\\
                             & > 110               & (exp.)  \cite{ANeumeier}\\

\hline
Rayleigh scattering          &90                   & (calc.) \cite{Rayleigh_th} \\
length (cm)                   
                             & 55 $\pm$ 5          &  (calc.) \cite{Rayleigh_th2}\\

\hline
\end{tabular}
\end{table}

\section{Details of the experiment }
\label{sec:setup}

\begin{figure}
\centering
  \includegraphics[width=0.8\linewidth]{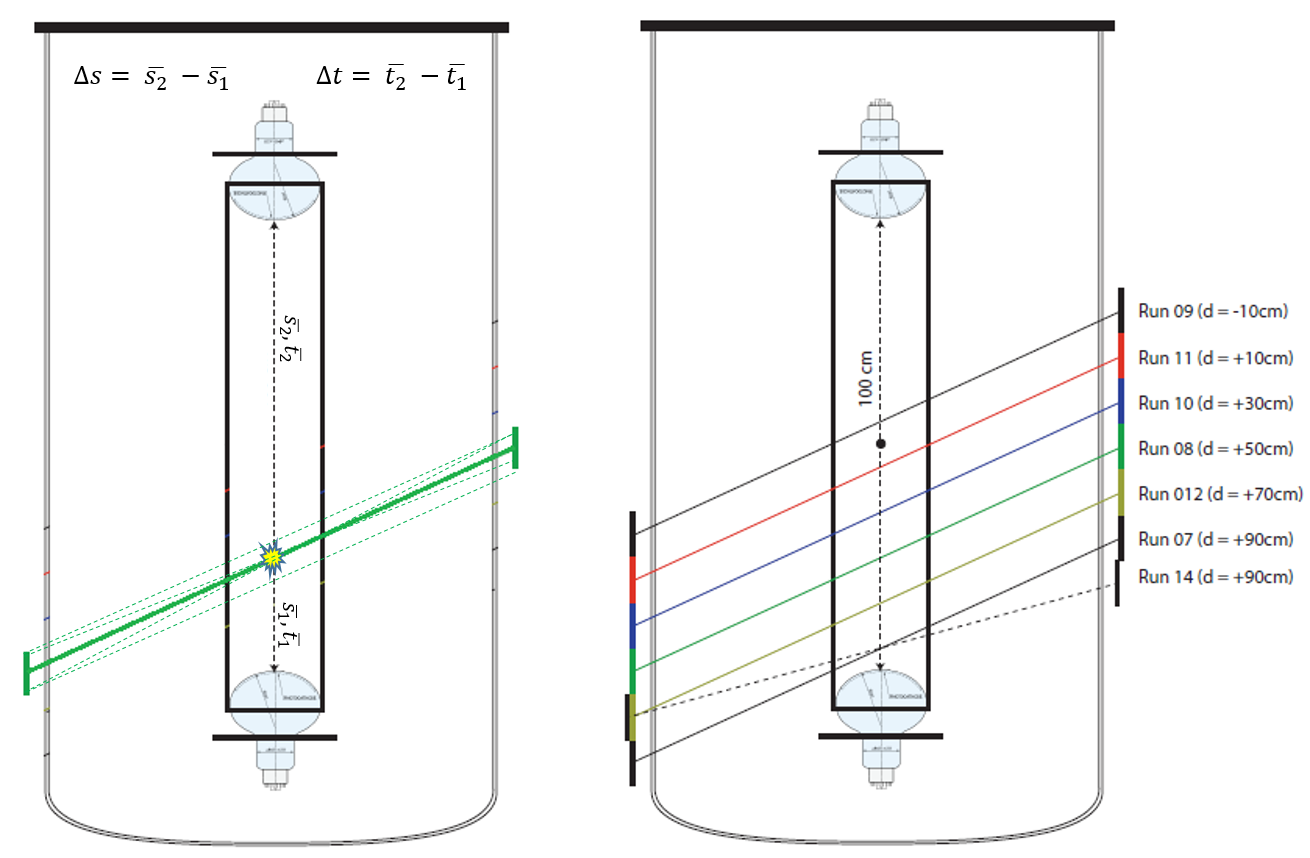} 
\caption{ \label{fig:expsetup} Schematic drawing of the experimental setup together with the movable trigger system. (Right) The two internal PMTs are located at the extremities of a box and immersed in liquid argon. Outside the cryostat, located symmetrically with respect to its diameter, two scintillator bars (colored lines) play the role of cosmic hodoscope and allow selecting cosmic ray tracks crossing the box at given distances with respect to the internal PMTs. For a given distance, different track slopes can also be selected (e.g. Run 07 and Run 14 in the  figure).\\
(Left) Sketch of how $\Delta s$ and $\Delta t$ are obtained in the measurement for a single position. Full discussion about the two quantities and their uncertainties can be found in Section~\ref{sec:Uncertainties}.
}
\end{figure}

The measurement of the velocity of LAr scintillation light presented in this paper is based on a simple setup. 
Two photomultiplier tubes (PMT) are immersed in LAr and positioned facing each other at a distance of 1~m. An external movable cosmic hodoscope is positioned symmetrically around the cryostat, allowing for triggering muons crossing the dewar at a variety of distances from the PMTs. By measuring the difference in path-lengths and of the light arrival time at the PMTs for a number of positions of the external hodoscope, the scintillation light velocity can be extracted from a simple linear fit. A schematic drawing of the experimental setup is shown in Figure~\ref{fig:expsetup}. A detailed description is presented in the next section.

\begin{figure}
 \centering
 \includegraphics[width=0.55\linewidth]{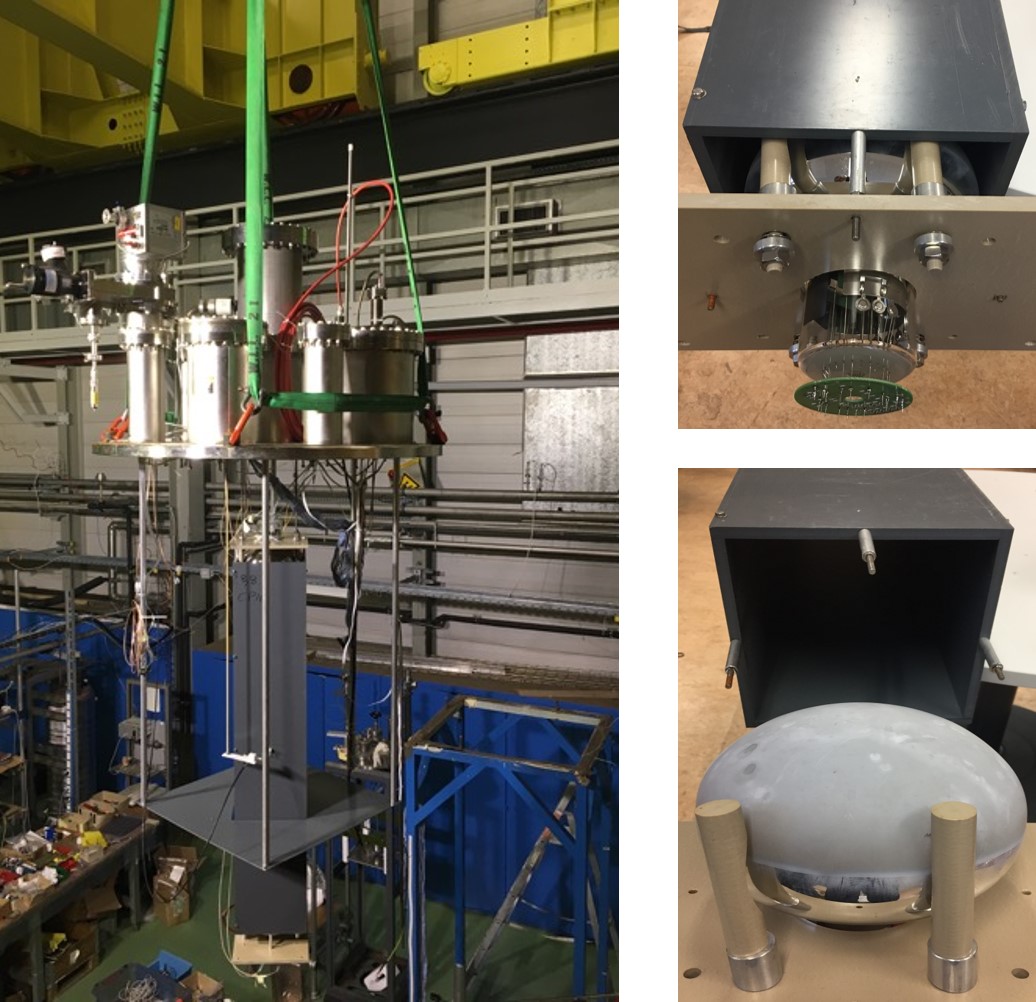} 
 \caption{ \label{fig:setup} Left: Experimental setup during the insertion in the liquid argon cryostat. Right: Picture of one internal PMT on its PEEK support (bottom) and in the final position inside the box (top).  
}
\end{figure}

\subsection{The setup}

The light detection system is made of two  8'' R5912-MOD, Hamamatsu 10-dynode photomultipliers, which are used to collect the LAr scintillation light.  The characteristics of the PMTs are reported in Table~\ref{tab:PMTsSpec} and in~\cite{QE,ICARUSPMTs,PMTsaturation}. 
In order to ensure Vacuum Ultra-Violet (VUV) light detection, the PMT windows were coated, by means of evaporation, with a layer of about $230\,\mu$g/cm$^2$ Tetraphenyl-butadiene (TPB), a wavelength shifter commonly employed in LAr-based rare-events experiments~\cite{GL2}. The shifted VUV light is re-emitted with a spectrum peaked at 430 nm, which is perfectly compatible with the collection spectrum of the chosen PMTs. Considering the devices own properties and the TPB conversion efficiency, the overall signal detection efficiency for the PMTs is estimated to be 12\%~\cite{GL2}.

\begin{table}
\centering
\caption{\label{tab:PMTsSpec} Summary of the R5912-MOD characteristics~\cite{QE, ICARUSPMTs} }
\begin{small}
\begin{tabular}{ll}
\hline
Spectral Response      & $300\div 650$~nm \\
Window Material        & Borosilicate (sand blasted) \\
Photocathode           & Bialkali with Pt under-layer   \\
Max suppy voltage & 2000 V \\
Photocathode Q.E. at 420nm  & $15 \div 16$\%  \\
Typical Gain           &  $1\times 10^7$ at 1500 V\\
Nominal P/V ratio$^*$ & 2.5 \\
Max. dark count rate$^*$   & 5000~s$^{-1}$ \\
Anode pulse rise time$^*$ & 3.8~ns\\
Electron transit time & 54~ns \\
Transit time spread & FWHM 2.4~ns\\
\hline
$^*$ Values for $G =1\times 10^7$ & \\
\end{tabular}
\end{small}
\end{table}



The experimental active volume is defined by a square box made of four vertical walls of hard, opaque polyethylene, 120~cm high. The two opposite ends of the box are occupied by the PMTs, each supported by a PEEK mechanical structure directly fixed on the box wall. The support is made in such a way that the main body of the PMT is inside the box walls, and the two internal PMTs fixed at a distance of about 100 cm. This defines a square-parallelepipedal active volume, 100 cm high and with a side of 21 cm. The position of the two PMTs with respect to the walls is defined in order to minimise light losses.

The active volume box and the PMTs are hanging from the main flange of the LAr cryostat as shown in Figure~\ref{fig:setup}. 
Next to the active volume box, a capacitive level meter, 30~cm long, is installed in order to monitor the level of the LAr and ensure it remains above the top PMT base. The main cryostat flange is equipped with different sizes UHV-quality feedthroughs to avoid liquid pollution, which include: 

\begin{itemize}
\item extraction of bias/signal cables for the two PMTs;
\item feedthrough for optical fiber and level meter;
\item connection to a vacuum pumping system (primary + turbomolecular pump);
\item liquid/gas input and gas exhaust;
\item a glass window to have a view of the inside during the filling.
\end{itemize}



An optical fiber is installed halfway between the two internal PMTs and optically coupled to a solid state 5~mW red laser diode, allowing for periodical PMTs stability checks in terms of gain and timing. The laser diode is controlled by a commercial Hewlett-Packard fast pulser providing light pulses with a duration of 3.8~ns FWHM and a rise time of 1.2~ns. The signal amplitude is regulated in order to have around 100 photoelectrons on each PMT for transit time measurement and of the order of a single electron for gain evaluation. In order to illuminate both PMTs with enough amount of light, orthogonal to the fiber, 1~mm of cladding is removed at the center of the fiber.



\subsection{Operations}

Commissioning of the setup starts by vacuum-pumping the double-walled volume of the LAr cryostat, down to a pressure of a few $10^{-3}$ mbar. The main volume is also isolated and vacuum pumped for a couple of hours, down to a few $10^{-4}$ mbar, to allow for outgassing. The dewar is first flushed with gaseous Argon-6.0\footnote{The notation 6.0, or 60 depending on the vendor, commercially refers to the purity of gases: in this case Argon-6.0 corresponds to liquid argon with 99.9999\% purity.}, and then filled with liquid Argon-6.0, without further purification. 
The liquid used for filling is certified to have a purity better than 1~ppm (O$_2$ equivalent). Given the dimensions of the experimental setup, the impurity level of Argon-6.0 ($\leq 0.5 \; ppm \; N_2$, $\leq 0.2 \; ppm \; O_2,\,H_2O,\,CO_2$) is expected to be low enough to allow for survival of a sufficient amount of scintillation light for the measurement~\cite{Acciarri_2010}. 

During standard operations, the system is operated in open loop, i.e. keeping it in over pressure with respect to the environment, and letting LAr evaporate. The exhaust line is opened right after filling is initiated; the evaporated argon is released outside the building via a long pipe, to avoid air back-diffusion and light leaks as well as to maintain the system in over pressure. All along the data-taking period the system is kept in over pressure between +0.1~bar and +0.3~bar. In these conditions the liquid argon temperature is between 88 K and 90 K~\cite{LArBook}.  


As mentioned above, the LAr level is set by means of a capacitive level meter, with a precision of around 2 mm. 
Heat injection through the dewar induces an evaporation rate of around 10-11~cm/day. A LAr re-fill is then done every 24~hours, in order to maintain the top PMT always immersed in the liquid. 


\subsection{The trigger and DAQ system}
\label{sec:exttrigger}

The trigger system is positioned outside the dewar and consists of two scintillating pads, 1~cm thick, 50~cm long and 10~cm wide, each one 
coupled via light guide to a PMT.  
The pads are mounted horizontally on support bars, fixed to the main flange in opposite positions with respect to the centre of the LAr dewar at an horizontal distance of 130~cm.  
The trigger pads supports allow for discrete placement along the height of the system within a few mm. This allow for selecting tracks at different heights with respect to the internal PMTs (see Figure~\ref{fig:expsetup}).

The actual trigger is formed by requiring a coincidence of both the two external and internal PMTs within a gate of 150~ns, which results in a rate of about 100 events per hour. For each triggered event, waveforms for both internal PMTs are recorded using an oscilloscope with a sampling rate of 5 GHz. The waveforms are recorded during a time window of 5 $\mu s$ to include most of the LAr scintillation light slow component. An interval of 0.5 $\mu s$ prior to the trigger is included in the recorded time window to allow for precise evaluation of the baseline. 


\section{Data Quality Monitoring}
\label{sec:DQM}

The data taking campaign took several weeks, during which the stability  of the experimental setup in terms of PMTs gain and LAr purity was constantly monitored.


\subsection{PMT monitoring}
\label{sec:PMTmonitoring}



The PMT characterization at LAr and room temperature is described in detail in~\cite{GL}. 
Once the dewar is filled with liquid argon and PMTs reach thermal stabilization, the high voltage is regulated in order to reach a gain of $~5\times10^6$ on each PMT. The gain is checked periodically, measuring the single electron response by means of a multichannel analyser.  No significant gain variation was observed during the whole data taking period. 

Several measurements~\cite{ICARUSPMTs} have shown that the transit time of PMTs does not change from ambient temperature to LAr/LN$_2$ temperature, once the high voltage is fixed. Periodical checks were done by comparing the arrival time measured on the internal PMTs, with respect to the trigger of the pulser. Any change observed is within the resolution of the instrument (200 ps).

It is worth reminding here that for the kind of measurement we want to perform, there is no need of any other specific inter-calibration as further clarified in Section~\ref{sec:analysis}. Therefore, any other PMT-associated cross-check is not needed for this specific measurement.

\subsection{LAr Purity}
\label{sec:LArPur}

As described in Section~\ref{sec:setup} the experimental setup does not include any purification system, since the measurement deals only with scintillation light and not with the ionization charge.

\begin{figure}
\centering
  \includegraphics[width=0.5\linewidth]{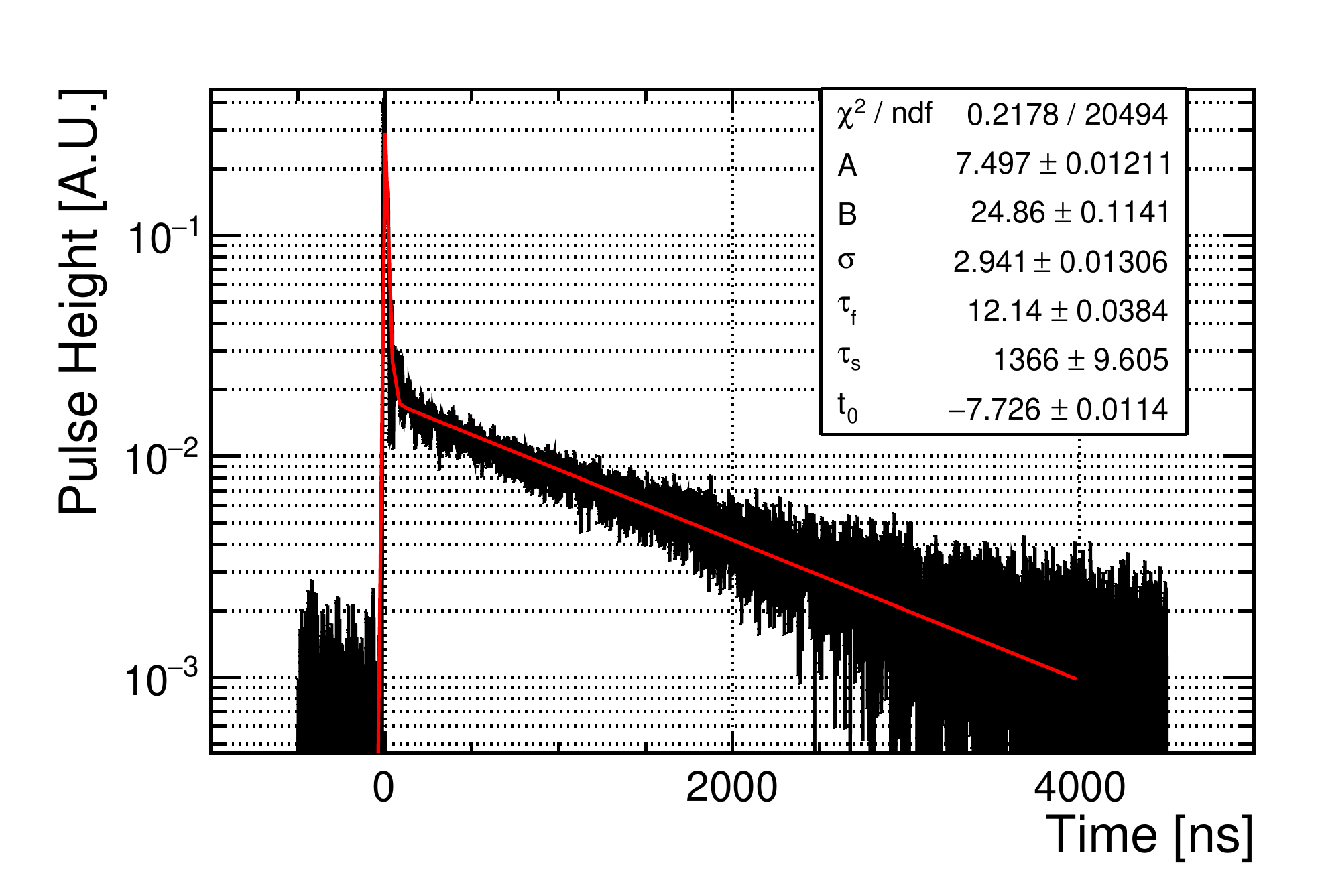} 
\caption{ \label{fig:FitTau3} Typical signal waveform from one of the two internal PMTs. The fit to the scintillation light slow component is also shown. }
\end{figure}

The level of air, $O_2, N_2, H_2O$ impurity concentrations in the liquid argon volume increases the probability of the argon dimer to undergo a non-radiative collision with the impurity molecules. Non-radiative quenching process is in competition with the de-excitation process of the long living triplet state. On the contrary the singlet state, because of its very fast decay time, is less sensitive to the presence of impurities~\cite{Mavrokoridis}. Consequently, since the arrival time of the scintillation light used in the analysis is extracted from the fast component, we consider the measurement robust against LAr purity degradation.
Nevertheless, in order to monitor the consistency of the overall dataset in terms of signal intensity, the purity of liquid argon was regularly checked by looking at the time constant for the slow component of the scintillation light. During the data-taking period the slow component was measured to lie within 1.4~$\mu s$ (evaporation phase) and 1.6~$\mu s$ (after re-fill). This range is compatible with an overall impurity concentration that is of the order of $< 1$~ppm, which translates into a negligible effect on the slow scintillation light component~\cite{Acciarri_2010}. Furthermore, the result is compatible with the certified Argon grade used in the experiment. Figure~\ref{fig:FitTau3} shows an example of signal waveform from one of the internal PMTs, together with the fit to the scintillation light slow component.

\section{Analysis strategy}
\label{sec:analysis}


In this section we describe the data samples used for the analysis as well as the method used to extract the difference of the photons arrival time .

\subsection{Data samples }

Six positions of the external trigger are considered for the measurement, thus defining six propagation path lengths. 
The external scintillator pads were positioned with a few mm precision in order to have cosmic tracks crossing, on average, the centre of the LAr active volume at the distances reported in Table~\ref{tab:dist}, with respect to the PMTs.

\begin{table}
\centering
\caption{\label{tab:dist} Summary of the data samples recorded. The first two columns show respectively the track mean distance from the bottom and the top PMTs. The third column gives the path-length which will be considered in the analysis and which corresponds to the difference between the first two columns.}
\begin{tabular}{c  c  c }
\hline
\hline
Mean distance from & Mean distance from & Path-length difference \\
bottom PMT [cm] & top PMT [cm] & d [cm] \\
\hline
5 & 95 & 90 \\
15 & 85 & 70 \\
25 & 75 & 50 \\
35 & 65 & 30 \\
45 & 55 & 10 \\
55 & 45 & -10 \\
\hline
\hline
\end{tabular}
\end{table}

To collect sufficient statistics to perform the analysis, (about 2500-3000 events), the data taking is repeated at each trigger position a few times following a random order. 
Data are collected for the entire set of positions for two different track slopes but with the same mean distance from the internal PMTs (see Figure~\ref{fig:expsetup}). 
The two track slopes considered are $\tan{\alpha} = 0.23$ and $\tan{\alpha} = 0.39$, corresponding to a vertical mutual distance of the pads of 30 and 50 cm, respectively.
The comparison of the results from these two independent samples of data allows for verification of the stability of the measurements with respect to the amount of light seen by each PMT.


All recorded data are considered for the analysis, except those events where the waveform from at least one of the two internal PMT was found to be out of the oscilloscope range.


\subsection{The constant fraction method}

The estimation of the light arrival time at the PMTs is done using a software constant fraction technique: on an event by event basis, the light arrival time is estimated as the time corresponding to a given percentage (e.g.  50\%) of the maximal signal amplitude. Once the signal arrival time at each PMT is extracted, the time difference is computed. 
Collecting data at different external trigger positions, one obtains two values ($\Delta t$, $\Delta s$) that can be put on a \textit{space vs time} plot. The light velocity can then be extracted as the slope of the straight line fitting the data points.

It is worth noticing here that the search for a given value of constant fraction is done both starting from the bottom or the top of the pulse, and only events returning the same time sample for both directions of search are considered. This simple criterion allows removing noisy events and consequently gives a stronger stability to the obtained results. 

The choice of the constant fraction method for the analysis is driven by the following consideration. Considering the time of the first photon arriving to the PMTs can lead to a bias in the measurement:  the \emph{real} first emitted photon  can get lost due to recombination, absorption or scattering on the walls. Furthermore, the first photon can also have a wavelength other than the mean value and since in this region the refractive index is expected to vary rapidly with wavelength, the velocity measurement would be biased. 
On the other hand, the large statistics of photons collected by the PMTs are likely to correspond to the median of the LAr scintillation wavelength distribution (e.g. $128 \pm 10$~nm as in Figure~\ref{fig:LArLightspectraA}).

To validate the assumption that the above described strategy gives access to the median wavelength of the scintillation light, a number of constant fraction values (40\%, 50\% and 60\%) are tested. For all of those central values, compatible results are observed.

\begin{figure}
 \centering
 {  \includegraphics[width=0.6\textwidth]{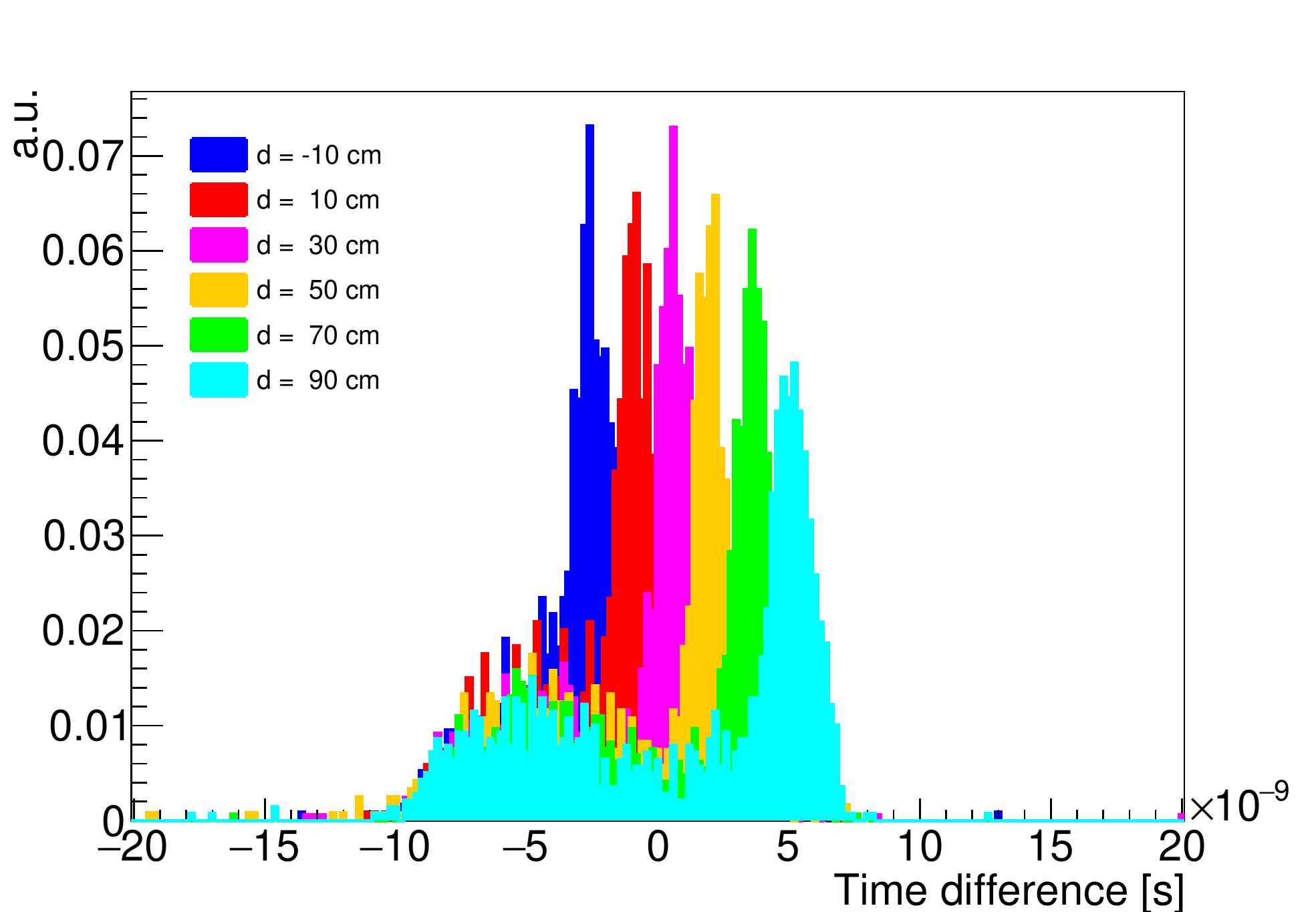} }
 \caption{\label{fig:timediff} Time difference between the top and the bottom PMT for the recorded events at different external trigger positions shown for a constant fraction of 50\% as an example.  }
 \end{figure}
 
 Figure~\ref{fig:timediff} shows the time difference distributions estimated at 50\% of the maximal amplitude, for the six external trigger positions and a given track slope. In each distribution the contribution of cosmic showers can be clearly disentangled from the Minimum Ionising Particle (MIP) events, for which the distribution is Gaussian with a mean depending on the external trigger position. Cosmic showers show in fact wide distributions always located at the same negative intervals, which are compatible with particles crossing the detector vertically, from the top to the bottom of the cryostat. From this distribution and the fit performed to take into account the contribution of the cosmic background below the distribution of interest, we consider the assumption of a stable background for all trigger configurations as correct. During the period of data collection the surrounding of the experimental setup did not change. 
 
The time difference values quoted in the next steps of the analysis are the mean and its associated error obtained from the Gaussian fit function representing the MIP contribution in Figure~\ref{fig:timediff}.





\begin{figure}
\centering
\begin{subfigure}{.4\textwidth}
  \centering
  \includegraphics[width=.9\linewidth]{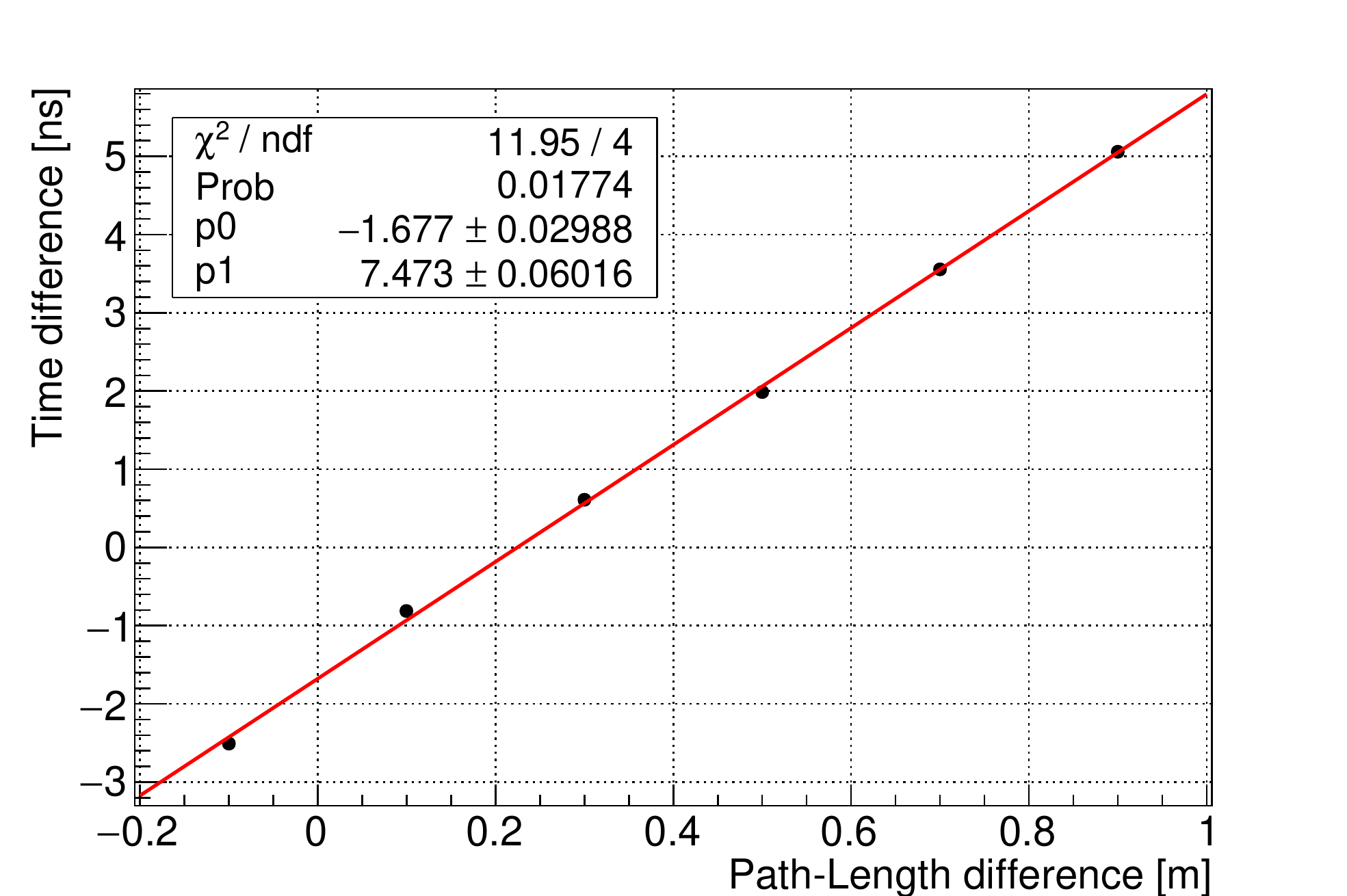}
  \caption{ \label{fig:gr_40pc_tg0p5} }
\end{subfigure}%
\begin{subfigure}{.4\textwidth}
  \centering
   \includegraphics[width=.9\linewidth]{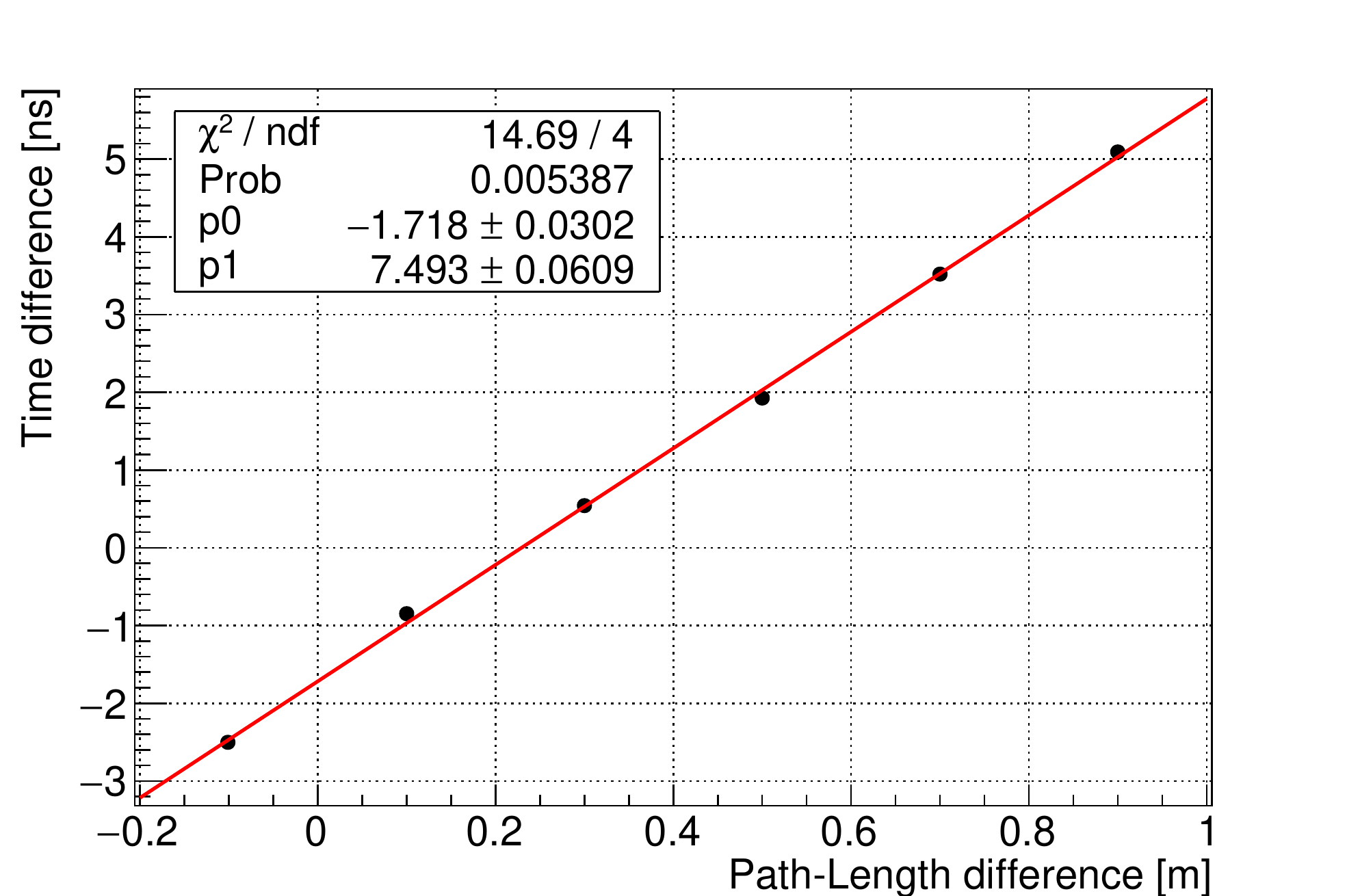}
  \caption{\label{fig:gr_50pc_tg0p5} }
  \end{subfigure}
\begin{subfigure}{.4\textwidth}
  \centering
  \includegraphics[width=.9\linewidth]{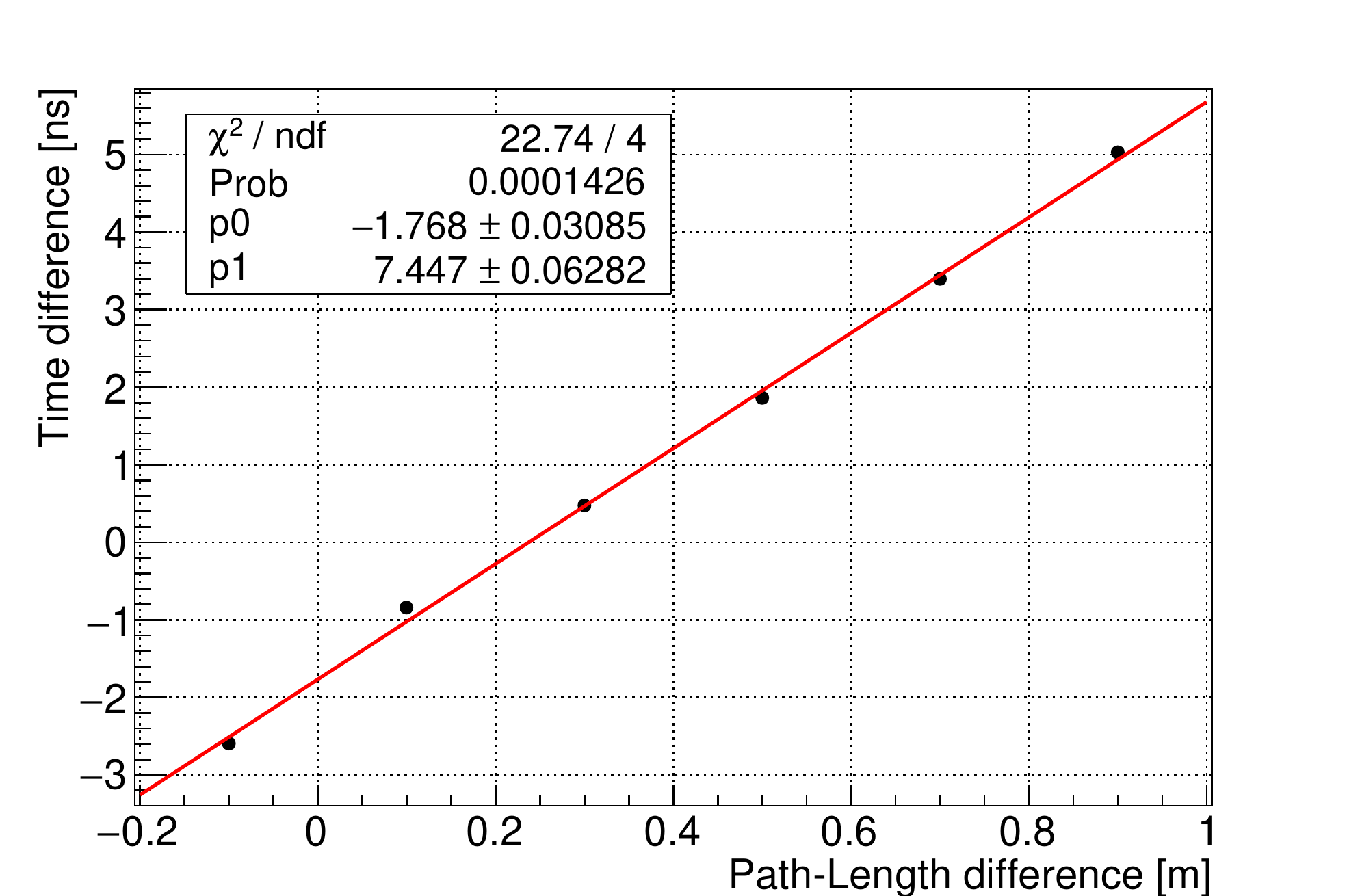}
  \caption{\label{fig:gr_60pc_tg0p5} }
  \end{subfigure}
\caption{\label{fig:gr_tg0p5} Examples of the extraction of the scintillation light velocity for three different constant fractions values, studied to estimate the light arrival time difference between the top and the bottom PMT. From left to right constant fraction at 40\%~(a), 50\%~(b) and 60\%~(c). The plots refer to the data samples taken with $\tan{\alpha}=0.23$. }
\end{figure}

Figure~\ref{fig:gr_tg0p5} shows the measurements obtained performing a linear fit of the time difference \textit{vs} distance data points, for a given track slope ($\tan\alpha=0.23$) taken as an example, for the three constant fraction values considered (40\%, 50\% and 60\%). 

A complete overview of the values of the inverse of the light velocity measured for each constant fraction considered and each track sample ($\tan\alpha=0.23$ and $\tan\alpha=0.39$) is presented in Table~\ref{tab:vg_results}. From this table we conclude that there is no dependency on the measurement from the constant fraction considered nor on the track slope. In fact, all measurements are compatible within the error reported, which takes into account the systematic uncertainties coming from the external trigger positioning. The other possible sources of systematic uncertainties are discussed in Section~\ref{sec:Uncertainties}. 




\begin{table}
\caption{\label{tab:vg_results} Summary of the results for the inverse of the scintillation light velocity in liquid argon, considering the two sets of tracks ($\tan\alpha=0.23$  and $\tan\alpha=0.39$) and a number of constant fraction values, namely 40\%, 50\% and 60\%. }
\centering
\begin{tabular}{ c c c }
\hline
\hline
 Constant fraction & Track sample 1 & Track sample 2 \\ 
                & $\tan\alpha=0.23$ & $\tan\alpha=0.39$ \\ 
                & [ns/m]              & [ns/m] \\    
 \hline

 40\% & 7.47 $\pm$ 0.06 & 7.46 $\pm$ 0.06 \\  
 50\% & 7.49 $\pm$ 0.06 & 7.54 $\pm$ 0.06 \\  
 60\% & 7.45 $\pm$ 0.06 & 7.48 $\pm$ 0.06 \\  

\hline
\hline
\end{tabular}

\end{table}

\begin{figure}
 \centering
 {  
 \includegraphics[width=0.5\textwidth]{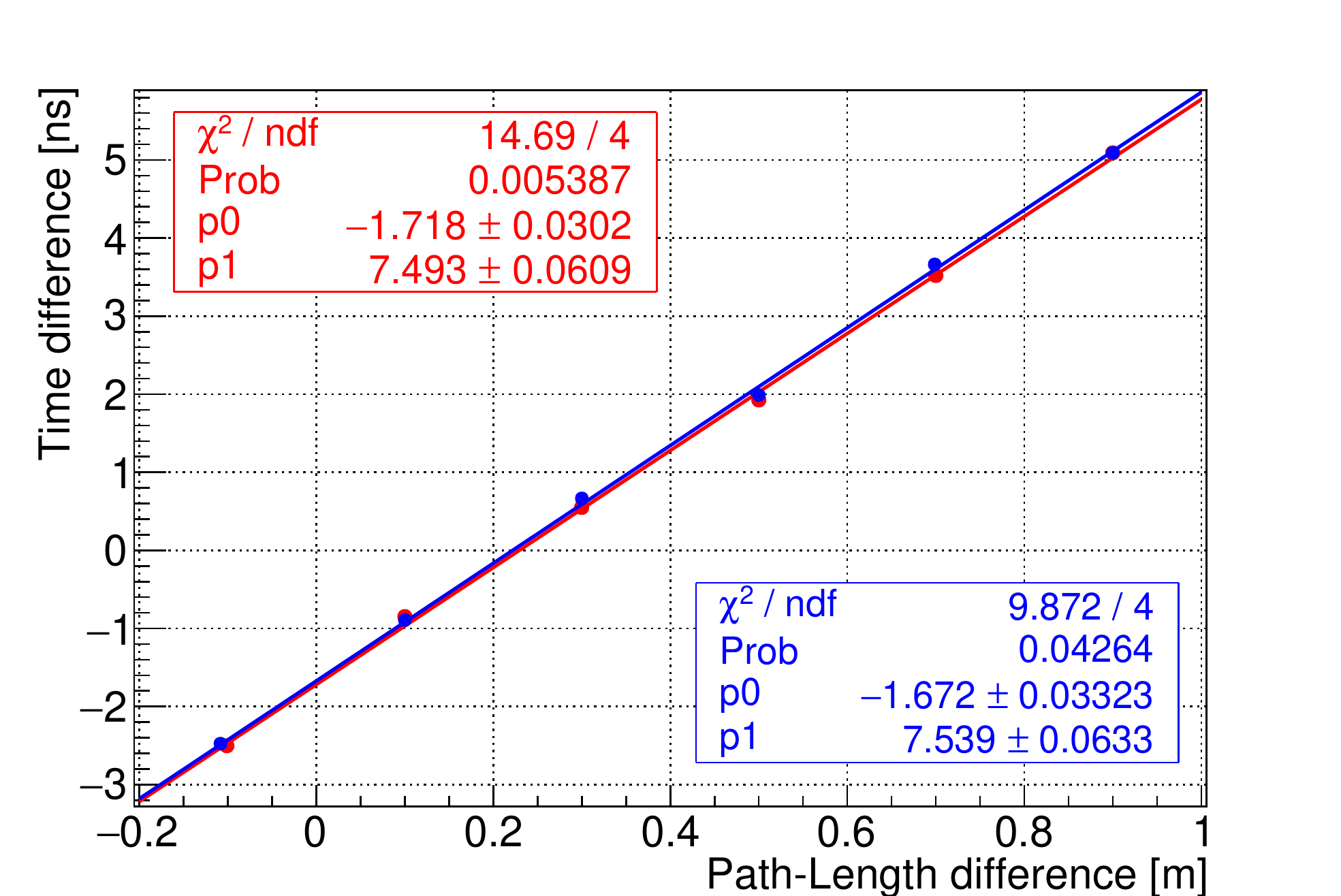}
 }
 \caption{\label{fig:SummaryGraph} Measurements of the scintillation light velocity in liquid argon performed with the two samples of tracks ($\tan{\alpha}=0.23$  and $\tan{\alpha}=0.39$). The measurements correspond to a constant fraction of 50\%.}
 \end{figure}



Figure~\ref{fig:SummaryGraph} shows the comparison of the measurement for the inverse of the scintillation light velocity in liquid argon for the two track slopes considered and a constant fraction of 50\%. The two extractions are in very good agreement.  

\subsection{Uncertainties on the measurement}
\label{sec:Uncertainties}

\subsubsection{Experimental uncertainties}

As explained in the previous section, this measurement relies on the estimation of two relative quantities, i.e. the difference in time arrival ($\Delta t$) of the scintillation photons at the two PMTs and the difference in the path ($\Delta s$) traveled by those very photons. Figure~\ref{fig:expsetup} (left) shows a schematic drawing of how those quantities are estimated for the measurement.  

For a given muon crossing path determined by the trigger system, one obtains a $\Delta s$ value as the difference between the two distances travelled by the photons to reach any of the two PMTs (see Table~\ref{tab:dist}). Moving the external hodoscope to a different position changes the relative difference in travelled distance by the photons (i.e. new value of $\Delta s$), but it does not affect any aspect of the inner setup. Therefore, any uncertainty related to the placement of the inner PMTs, affecting a single $\Delta s$ estimation, carries over to the new data point unchanged.
The same reasoning is valid when considering the difference in photon travel time ($\Delta t$), which on a single measurement would depend on the actual distance difference covered by the photons, as well as on intrinsic characteristics of the PMTs, like transit time and TPB conversion time. When moving the hodoscope to a new position, the PMTs properties remain the same, and the change in $\Delta t$ is only affected by the corresponding variation of $\Delta s$.

When taking multiple measurements, all mentioned inner PMT-related uncertainties can be treated as a constant offset affecting all data points. This offset can be estimated experimentally, but it is completely irrelevant to the actual measurement of the physical quantity of interest here, which is extracted from the slope of the linear relation between $\Delta t$ and $\Delta s$. As a result,
the only uncertainty affecting the analysis is related to the positioning of the scintillating pads of the trigger system. 
In the analysis, an error of 0.5~cm is considered, given by the achievable precision in the positioning of the external PMTs mechanical support.


Another source of uncertainty related to the trigger system arises from the large geometrical cross-section of the pads, $50 \times 10$ cm$^2$, which translates into a wide solid angle over which cosmic tracks can be distributed. This is actually mitigated by collecting and measuring the time difference from a large population of events, whose mean path is well defined from the Gaussian shape of the distribution.
It can be further noted that the stability of the obtained velocity result, against the change in the track slope selected through the hodoscope (see Table~\ref{tab:vg_results}), demonstrates that such uncertainty is mitigated. 

\subsubsection{Uncertainties derived from Monte Carlo simulation}
\label{sec:simulation}

Simulation studies were performed with FLUKA~\cite{FLUKA1,FLUKA2} to probe the consistency of the analysis strategy as well as evaluate possible additional sources of systematic uncertainties. 
The PMTs geometry is reproduced in FLUKA according to the technical drawings and their location is set at 1~m distance from each others. The simulation also includes external scintillators with real dimensions and adjustable vertical position. For each simulated event a monochromatic 5 GeV/c muons is generated next to the top scintillator counter and directed towards the bottom one. 
The muons starting positions are randomly chosen over an area covering the top scintillator, and an angular divergence of 100~mrad is included. 
The total yield of scintillation photons is set to $5.1\times 10^4$ phe/MeV as in~\cite{Doke}, and the scintillation light spectrum is simulated with a Gaussian shape centered at $\lambda = 126.8$~nm and a FWHM~$=7.8$~nm~\cite{light_production_exp}. 
The emission of Cherenkov light and of the visible light produced by the TPB are also simulated. LAr is assumed to be fully transparent to its own scintillation light. The  Rayleigh Scattering length is set to 90~cm~\cite{Rayleigh_th}. 
The propagation speed is calculated according to the sampled photon wavelength. As a starting point, the parameterization of the refractive index given in~\cite{refIndex} is used, and the group velocity is calculated as :

 \begin{equation}
\label{eq:vg3}
v_g = \frac{c}{n -\lambda \frac{d
      n}{d \lambda} }
\end{equation}

\noindent $\lambda$ being the vacuum wavelength and $n$ the refractive index.

The arrival time of each optical photon on the PMT surface was scored and convoluted with a triangular shape signal with a rise time of 3~ns and an arbitrary gain factor of 10.

The resulting waveforms are analyzed with the same constant fraction method as applied to data, namely 40\%, 50\% and 60\% as well as for the two track slopes analysed in data.
In order to assess the sensitivity of the measurement, simulations are repeated with different values of the refractive index, meaning that the refractive index curve is rigidly shifted by $\pm 10$\% with respect to the default value of 6.46 ns/m. This implies having simulations for different group velocity inputs, namely:  5.82 ns/m and 7.11 ns/m. Finally, a dedicated study to quantify the possible error in our velocity estimation introduced by the Cherenkov effect has is performed.

\begin{figure}
    \centering
      \includegraphics[width=.5\textwidth]{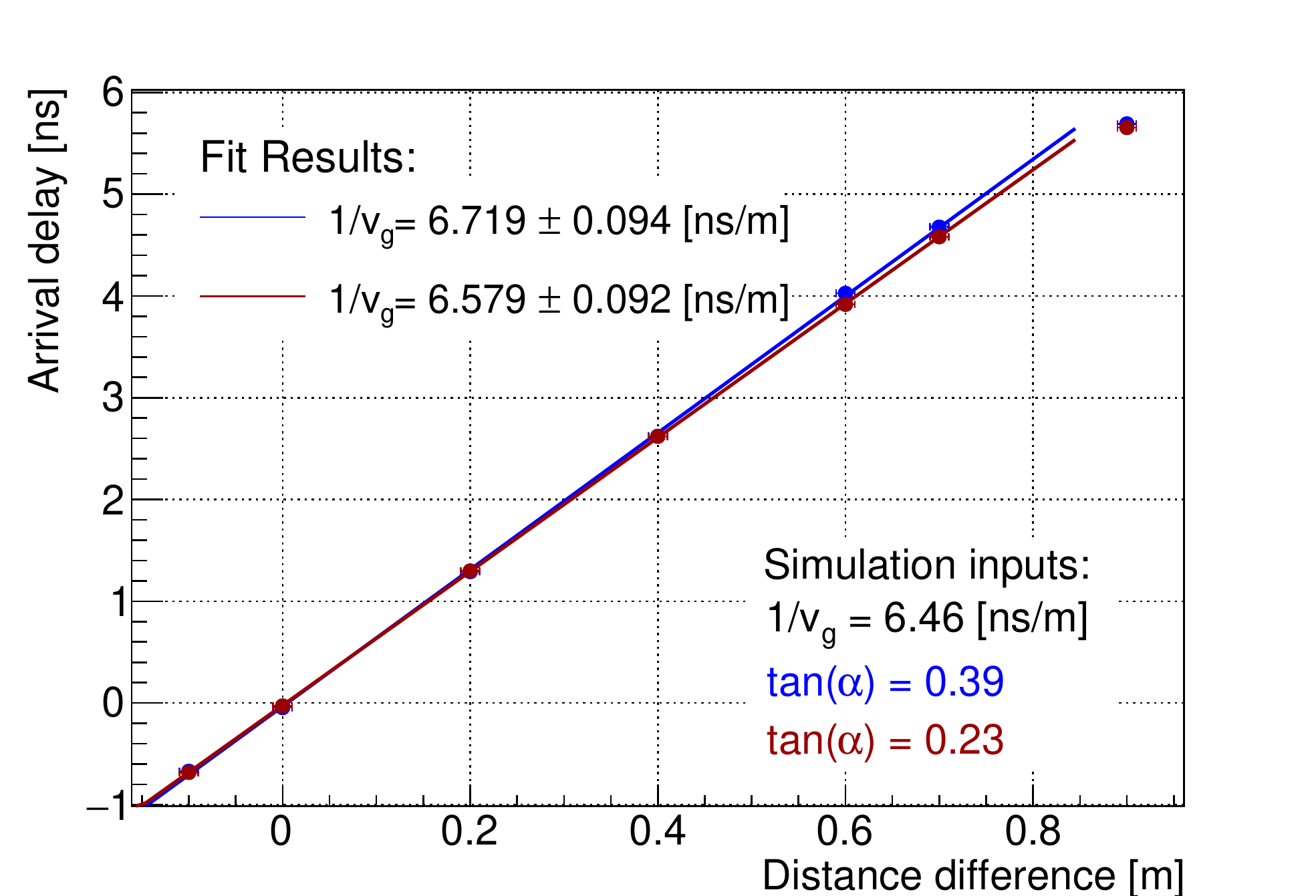}
      \caption{\label{fig:FLUKA_Cherenkov2} FLUKA simulation for $\frac{1}{v_g}= $ 6.46 ns/m for the two track slopes considered in the analysis ($\tan\alpha= 0.23$ and $0.39$) and a constant fraction at 50\%.  }
    \end{figure}

Table~\ref{tab:FLUKAvgvary} shows the stability of the velocity extraction with respect to the constant fraction considered. The initial assumption of the true value of the velocity is also tested. 
From these studies we derive that the extractions of the inverse of the light velocity in LAr is consistent with the true input value within 1.5$\sigma$. This result confirms the reliability of the constant fraction method applied on the data.  The values presented in Table~\ref{tab:FLUKAvgvary} correspond to the track slope of $\tan\alpha = 0.23$. However, similar conclusions can be derived by the other sample of simulated data which reproduce the other track slope. Therefore, we can conclude that there is no impact on the derived velocity coming from the trigger layout considered, which is in agreement with what we observe in data. As an example, Figure~\ref{fig:FLUKA_Cherenkov2} shows the agreement of the inverse velocity measurement obtained with FLUKA for the two trigger layouts considered for $\frac{1}{v_g}=6.46$~ns/m as an input value in the simulation. 

Finally, we studied the velocity measurement uncertainties due to both the Cherenkov light and the visible light produced by the TPB deposited on the surface of the PMTs. These effects might generate biases on the time difference estimation with sizes expected to vary with the position of the external trigger: 

\begin{itemize}
    \item the Cherenkov light can reach the PMT closer to the track before the UV light, thus increasing the arrival time difference between the two PMT's;
    \item the visible light produced on the surface on the PMT closer to the track can reach the PMT further away before the UV light, thus reducing the arrival time difference between the two PMTs.
\end{itemize}

The simulation shows that only the last point at d=90~cm might suffer from the mentioned effects in a sizeable way.
This explains the mis-alignement of that point observed in Figure~\ref{fig:FLUKA_Cherenkov2}. The quantification of the shift in time introduced by the Cherenkov and the visible light strongly depends on the assumptions made in the simulation. For that reason a very conservative systematic uncertainty of $\pm 1$~ns is evaluated for the d=90~cm point.
The velocity measurements extracted under these consideration are found fully compatible to the result previously presented. This indicates that, for the specific layout of our experimental setup, the Cherenkov and visible light effects compensate each other.

\begin{table}
\caption{\label{tab:FLUKAvgvary} Summary of the results obtained from the FLUKA simulation for the track sample with $\tan(\alpha) =0.23$ and the number of constant fraction considered in the analysis constant fraction (CF).  }
\centering
\begin{tabular}{cccc}
\hline
\hline
Simulation input  &  Extracted value       & Extracted value &  Extracted value \\
                  &   40\%                 &  50\%           &     60\%         \\
$1/v_g$~[ns/m]    & $1/v_g$~[ns/m]         &  $1/v_g$~[ns/m] & $1/v_g$~[ns/m]   \\ 
\hline

\hline
 5.82            &  $5.875 \pm 0.082$     &    $5.895 \pm 0.082 $   &  $5.923 \pm 0.083$   \\
\hline
6.46            &     $6.543\pm 0.092$      & $6.579 \pm 0.092$                  &   $6.602 \pm 0.093$ \\  
\hline
 7.11            &    $7.271 \pm 0.101$       &    $7.287 \pm 0.102   $           &   $ 7.315 \pm 0.102$ \\
\hline
\hline
\end{tabular}
\end{table}



\section{Results}
\label{sec:finresults}

\subsection{Measurement of the scintillation light velocity in LAr }

This section presents the results obtained for the measurement of the velocity of the scintillation light in LAr, taking into account the simulation studies.  
These simulations have shown that the data point having the external trigger positioned close to the bottom PMT (i.e. Path-length=~90~cm) is affected by large uncertainties. 
For that reason, the extraction of the final velocity measurement is performed excluding that point from the linear fit and the new values are reported in Table~\ref{tab:vg_results_no90cm}. These results are found to be fully compatible with those those previously shown in Table~\ref{tab:vg_results}.\\


The final value for the inverse of the velocity of the scintillation light in liquid argon is then $7.46 \pm 0.03 \mathrm{(stat)} \pm 0.07 \mathrm{(syst)}$~ns/m where the systematic error is computed as the standard deviation of the values reported in Table~\ref{tab:vg_results}. 

\begin{table}
\caption{\label{tab:vg_results_no90cm} Summary of the results for the inverse of the scintillation light velocity in liquid argon excluding the point at d~=~90~cm, to avoid any possible bias due to Cherenkov light as shown by the simulation. Numbers are quoted for the two sets of tracks ($\tan\alpha=0.23$  and $\tan\alpha=0.39$) and the constant fraction considered in the analysis ( 40\%, 50\% and 60\%). }
\centering
\begin{tabular}{ c c c }
\hline
\hline
 Constant fraction & Track sample 1 & Track sample 2 \\ 
                & $\tan\alpha=0.23$ & $\tan\alpha=0.39$ \\ 
                & [ns/m]              & [ns/m] \\    
 \hline
 40\% & 7.46 $\pm$ 0.08 & 7.53 $\pm$ 0.08 \\  
 50\% & 7.42 $\pm$ 0.08 & 7.57 $\pm$ 0.08 \\  
 60\% & 7.35 $\pm$ 0.08 & 7.42 $\pm$ 0.08 \\  

\hline
\hline
\end{tabular}
\end{table}

\subsection{Other derived quantities: Refractive index and Rayleigh Scattering length }
\label{sec:corollario}

The group velocity is related to the refractive index through its derivative as shown by Equation~\ref{eq:vg3}. The measurement performed can thus be used to extract the refractive index $n$ of liquid argon in the VUV region and in particular at 128~nm.
In a dispersive medium, the refractive index $n$ depends on the frequency $\nu$, as described by the Lorentz-Lorenz equation (\cite{Bornbook}):

\begin{equation}
\label{eq:n}
\frac{n^2 -1}{n^2 + 2} = \frac{4 \pi N \alpha}{3} = \frac{1}{3} \Sigma_{k} \frac{\rho_{k}}{\nu^2_{k} - \nu^2}
\end{equation}
where $N$ is the number of molecules per unit volume and $\alpha$ is the mean polarizability of the medium. Equation~\ref{eq:n} explicitly shows the coexistence of a number of resonance frequencies $\nu_k$ in the same system even if composed by molecules of the same kind. The coefficients $\rho_k$ are proportional to the number density and $N$ and the relative strengths
of the resonances.

The index of refraction can then be expressed as:

\begin{equation}
\label{eq:n2}
n = \sqrt{1 + \frac{3 x}{3-x}}
\end{equation}

where: 
\begin{equation}
\label{eq:x}
x = 4\pi N \alpha = \Sigma_{k} \frac{\rho_{k}}{\nu^2_{k} - \nu^2}
\end{equation}


The liquid argon resonances closest to the wavelength of interest (128 nm) are two:  the one in the UV region (106.6~nm) and one in the IR region (908.3~nm). Therefore Equation~\ref{eq:x} can be re-written as :

\begin{equation}
\label{eq:x2}
x = a + \Sigma_{k} \frac{b_k}{\nu^2 - \nu^2_k} =
a_{0} + \frac{a_{UV} \lambda^{2} }{ \lambda^{2}  -  \lambda^{2}_{UV}} + \frac{a_{IR} \lambda^{2} }{ \lambda^{2}  -  \lambda^{2}_{IR}}
\end{equation}
where the parameters $a_{0}$ , $a_{UV}$ and $a_{IR}$ known as \textit{Sellmeier coefficients} have to be determined experimentally.  \\


It is true that other resonances exist for liquid argon (e.g. 104.8 nm or 90 nm) and also for impurities such as O$_2$ and N$_2$. However to allow for a meaningful description of such resonances,  experimental datasets in the same energy range would be needed. Therefore given the available data in liquid, the most reliable parametrisation is the one given in~ \ref{eq:x2} where all other terms not explicitly appearing are summed up in the coefficient $a_0$. This is even more true for the resonances related to impurities, as their effect would be suppressed by a factor identical to their concentration factor, i.e. $10^{-6}$ at best.

Equations~\ref{eq:n2}, \ref{eq:x2} together with Equation~\ref{eq:vg3} are  used to derive the index of refraction of liquid argon in the VUV region.
The approach is similar to previous works~\cite{Hitachi,Rayleigh_th2}. A major difference is that, in this work, instead of using the gas approximation ($n \sim 1$), the full parametrisation of $n$ is used. In Appendix~\ref{sec:appendix} a comparison of the results obtained with the two parametrisations is discussed.

\begin{figure}
 \centering
 {  
 \includegraphics[width=0.9\textwidth]{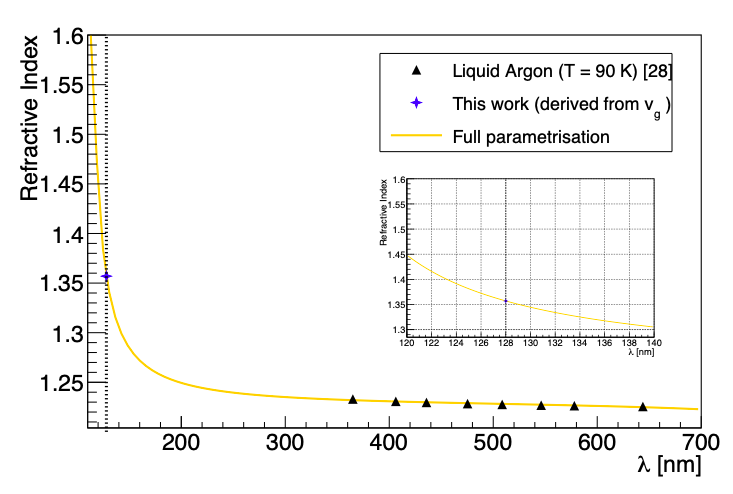} 
 
 }
 \caption{Extraction of the refractive index in the UV region and in particular for the typical wavelength of the scintillation light of LAr. The small plot shows a zoom-in of the region of interest. The 128~nm wavelength is pointed out by the dotted line. }
 \label{fig:RefIndexExtraction} 
 \end{figure}

Data recorded at 90~K for wavelength between 350 and 650~nm 
~\cite{PhysRev.181.1297} and our group velocity measurement are considered via a $\chi^2$ method to constrain the region of interest where $n$ varies rapidly.    
The curve obtained is shown in Figure~\ref{fig:RefIndexExtraction}. The inner box shows a zoom-in of the region between 110 and 140~nm. The resulting refractive index at 128~nm is $1.358 \pm 0.003$.
The values for the Sellmeier coefficients $a_{0}, \, a_{UV} \, \mathrm{and} \, a_{IR}$ are reported in Table~\ref{tab:RefIndexParam}. \\

\begin{table}
\caption{\label{tab:RefIndexParam} Sellmeier coefficients for liquid argon at 90~K extracted using data wavelength between 350 and 650~nm 
~\cite{PhysRev.181.1297} and this paper group velocity measurement to constrain the VUV region. }
\centering
\begin{tabular}{ c c c }
\hline
\hline
$a_{0}$ & $a_{UV}$ & $a_{IR}$ \\ 
\hline
 $0.334 \pm 0.008$ & $ 0.100 \pm 0.007 $   & $ 0.008 \pm 0.003$ \\  
\hline
\end{tabular}

\end{table}

Once the refractive index is extracted, the Rayleigh Scattering typical length can be derived~\cite{Landau, Rayleigh_th}:

\begin{equation}
\mathcal{L}^{-1} = \frac{16\pi^{3}}{6\lambda^{4} } \Big[ k T \kappa_{T} \Big( \frac{(n^{2}_{\lambda} - 1	) (n^{2}_{\lambda} + 2 )}{3} \Big)^2  \Big]
\label{RayScat}
\end{equation}
where $\mathcal{L}$ is the scattering length, $\lambda$ is the wavelength, $k$ is the Boltzman constant, $T$ is temperature, 
$\kappa_{T}$ the isothermal compressibility, and $n$ is the index of refraction corresponding the wavelength considered. 
For a temperature of 90 K and an isothermal compressibility of $2.21 \cdot 10^{-4} \mathrm{cm^{2}/kg}$  \cite{Jain_1971}, a scattering length $\mathcal{L}$ of $99.1 \pm 2.3$~cm is derived for   $\lambda=$128~nm.


\section{Conclusions}
\label{sec:conclusions}

In this paper we present a first direct measurement of the light group velocity in liquid argon. The measurement yields $\frac{1}{v_g} = 7.46 \pm 0.08$ ns/m. An estimation of $n$ and $\mathcal{L}$ is derived, considering the full parametrisation of $n$ for liquid media, yielding respectively $n= 1.358 \pm 0.003$ and $\mathcal{L}= (99.1 \pm 2.3$)~cm. 
A discussion of the derived quantities using the more common gas approximation is presented in the Appendix. 

\section{Acknowledgements}
\label{sec:thanks}

The authors wish to thank the CERN Neutrino Platform for providing the necessary infrastructures to perform the measurement and Alfredo Ferrari for the fruitful discussions and the help on the simulation.


\bibliographystyle{JHEP}
\bibliography{biblio}

\clearpage

\appendix
\section{Refractive index extraction with gas approximation}
\label{sec:appendix}

The expression of the refractive index (see Equation~\ref{eq:n}) can be simplified in case of gases, since $n$ is close to unity. In those cases the expression becomes:  

\begin{equation}
\label{App:eq:n}
n^2 -1 = 4\pi N \alpha = \Sigma_{k} \frac{\rho_{k}}{\nu^2_{k} - \nu^2}
\end{equation}

Following the same procedure as before, by considering only the resonances closest to the scintillation wavelength of argon this expression can be re-written as:      

\begin{equation}
n^{2} = a + \Sigma_{k} \frac{b_k}{\lambda^2 - \lambda^2_k} = a_{0} + \frac{a_{UV} \lambda^{2} }{ \lambda^{2}  -  \lambda^{2}_{UV}} + \frac{a_{IR} \lambda^{2} }{ \lambda^{2}  -  \lambda^{2}_{IR}}
\end{equation}
where the parameters $a_{0}$ , $a_{UV}$ and $a_{IR}$ have to be determined experimentally. \\

\begin{figure} [htbp]
 \centering
 {\includegraphics[width=0.9\textwidth]{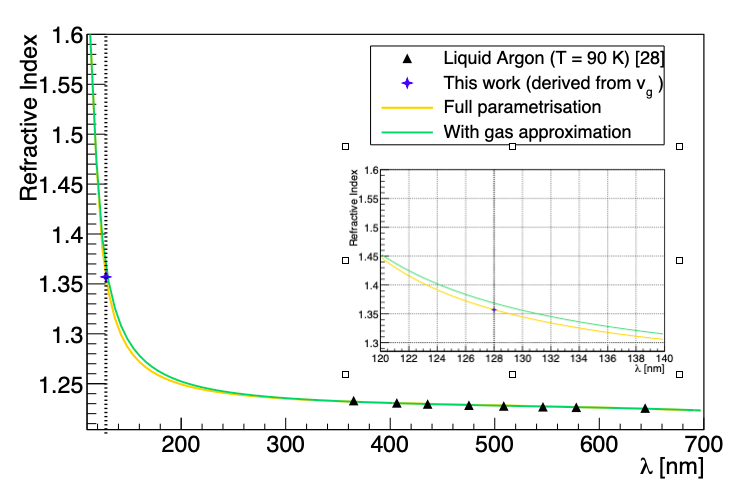}}
 \caption{\label{fig:RefIndexcomparison} Comparison of the refractive index extraction in the UV region using the full approximation (default results) or the gas approximation. The small plot shows a zoom-in of the region typical for the wavelength of the scintillation light of LAr. The 128~nm wavelength is pointed out by the dotted line. }
 \end{figure}

In the literature, sometimes the use of this approximation can be found also for liquid argon, although the approximation $n \sim 1$ is not fully justified. 
In Table~\ref{tab:RefIndexParamApp} we present the comparison of the refractive index value extracted using the approximated expression (see Equation~\ref{App:eq:n}) to the one obtained with the regular expression, which represents the nominal result. The values for parameters $a_{0}$ , $a_{UV}$ and $a_{IR}$ extracted from the fit are also reported. 
Figure~\ref{fig:RefIndexcomparison} shows graphically the comparison between the two extraction curves of the refractive index $n$. In this figure, it can be noticed that the main difference between the two parameterisation is in the region of $\lambda < 200 $~nm. 

To conclude this section we propagate this comparison to the value for the Rayleigh scattering length. When the gas approximation is used to extract the refractive index, the typical Rayleigh Scattering length is estimated at  $\mathcal{L} = 91.0 \pm 2.1$~cm  which corresponds to about 8\% less than the estimated value presented in the text.

\begin{table}[htbp]
\caption{\label{tab:RefIndexParamApp} Comparison of the extraction of the refractive index $n$ using the formula with gas approximation to the default result. The values for parameters $a_{0}$ , $a_{UV}$ and $a_{IR}$ extracted from the fit are also reported.   }
\centering
\begin{tabular}{c c c c c }
\hline
\hline
& $n$ & $a_{0}$ & $a_{UV}$ & $a_{IR}$ \\ 
\hline
Full expression & $1.358 \pm 0.003$ & $0.334 \pm 0.008$ & $0.100 \pm 0.007 $   &  $ 0.008 \pm 0.003$ \\  
With gas approximation & $1.369 \pm 0.003$ & $1.341 \pm 0.015$ & $ 0.163 \pm 0.014 $   & $ 0.008 \pm 0.004$ \\  
\hline
\hline
\end{tabular}
\end{table}

\end{document}